\documentclass[3p,times]{elsarticle}

\usepackage{ecrc,amssymb}
\usepackage[figuresright]{rotating}

\volume{00}
\firstpage{1}
\journalname{Nuclear Physics A}
\runauth{A. Ficnar et al.}
\jid{npa}
\jnltitlelogo{Nuclear Physics A}
\biboptions{square,comma,numbers,sort&compress}

\newcommand{\eqn}[2] {\begin{equation}\label{#1}{#2}\end{equation}}
\newcommand{\eno}[1] {(\ref{#1})}

\begin{document}

\begin{frontmatter}
\title{Holographic light quark jet quenching at RHIC and LHC \\ via the shooting strings}
\author[CU]{Andrej Ficnar}
\author[PU]{Steven S. Gubser}
\author[CU]{Miklos Gyulassy}
\address[CU]{Department of Physics, Columbia University, New York, NY 10027, USA}
\address[PU]{Joseph Henry Laboratories, Princeton University, Princeton, NJ 08544, USA}
\begin{abstract}
A new shooting string holographic model of jet quenching of light quarks in strongly coupled plasmas is presented to overcome the phenomenological incompatibilities of previous falling string holographic scenarios that emerged when confronted with the recent LHC data. This model is based on strings with finite momentum endpoints that start close to the horizon and lose energy as they approach the boundary. This framework is applied to compute the nuclear modification factor $R_{AA}$ of light hadrons at RHIC and LHC, showing that this model improves greatly the comparison with the recent light hadron suppression data. The effects of the Gauss-Bonnet quadratic curvature corrections to the $AdS_5$ geometry further improve the agreement with the data.
\end{abstract}
\begin{keyword}
gauge/string duality \sep quark-gluon plasma \sep energy loss
\end{keyword}
\end{frontmatter}


\section{Introduction}
\label{intro}

Light quarks in AdS/CFT can be modeled by strings with one endpoint ending on a D7-brane in the bulk of $AdS_5$. A possible way to model a light quark-anti quark pair that has undergone a hard scattering is by considering an initially pointlike open string created close to the boundary with endpoints that are free to fly apart and fall towards the black hole, the so-called ``falling'' strings \cite{ggpr08,Chesler:2008wd}. It was found \cite{ggpr08,Hatta:2008tx,Chesler:2008wd,cjky08} that the maximum distance such an energetic quark can travel for a fixed energy $E$ in a thermal $\mathcal{N}=4$ SYM plasma at a temperature $T$ scales as $\Delta x_{\rm max}\propto E^{1/3}T^{-4/3}$, where the constant of proportionality is important for phenomenological applications as it determines the overall strength of jet quenching. 

In order to use this model and compute the observables such as the nuclear modification factor $R_{AA}$, one needs to know the details of how these light quarks lose energy. The application of a general formula for energy loss in non-stationary string configurations \cite{f12} to the case of fallings strings has shown a seemingly linear path dependence in the phenomenologically relevant range. Simple constructions of $R_{AA}$ \cite{fng12} resulted in a serious under-prediction of the LHC pion suppression data (although it had the right qualitative structure), even with the inclusion of the higher derivative corrections to the $AdS_5$, indicating that the predicted jet quenching was too strong. 

A possible resolution of this problem was recently offered by considering strings that have finite momentum at their endpoints \cite{fg13}, as the stopping distance of light quarks dual to these strings is greater than in the previous treatments of the falling strings, and hence may offer a better match with the experimental data. Furthermore, these strings provide a more natural holographic dual of dressed energetic quarks, as one can think of the finite momentum endpoints representing quarks themselves and the string between them the color field they generate. In this way, one also obtains a clear distinction between the energy in the hard probe and energy contained in the color fields surrounding it, thus offering a precise definition of the instantaneous jet energy loss that was missing in earlier accounts.


\section{Energy loss}
\label{enloss}

Endpoints with finite momenta move along null geodesics in $AdS_5$-Schwarzschild and the evolution of their momenta is governed by equations that do not depend on the bulk shape of the string, only on the radial position of the endpoint \cite{fg13}:
\eqn{eloss1}
{\frac{dE}{dx}=-\frac{\sqrt{\lambda}}{2\pi}\frac{\sqrt{f(z_*)}}{z^2}\,,}
where $\sqrt{\lambda}=L^2/\alpha'$ is the 't Hooft coupling, $f(z)=1-z^4/z_H^4$ (the boundary is at $z=0$), $z_H=1/(\pi T)$ and $z_*$ is the minimal (inverse) radial coordinate the geodesic reaches and which hence completely determines the motion of the endpoint. As mentioned before, the finite endpoint momentum provides a very natural definition of the quark energy loss as precisely the rate \eno{eloss1} at which the energy gets drained from the endpoint. 

For phenomenological purposes, we need to express \eno{eloss1} as a function of $x$, which means that we need to solve the null geodesic equation. If, initially, at $x=0$, the endpoint is located at $z=z_0$ and is going towards the boundary, the solution to the geodesic equation has a strongly convergent expansion for small $z_*$:
\eqn{eloss3}
{x_{\rm geo}(z)=z_H^2 \left[ \left(\frac{1}{z}-\frac{1}{z_0}\right)+\mathcal{O}\left(\frac{z_*^4}{10 z^5},\frac{z_*^4}{10 z_0^5}\right) \right] \,.}
This expansion is interesting for phenomenological reasons: we see that for small $z$ the energy loss \eno{eloss1} is large and therefore quarks dual to endpoints that move close to the boundary will be quenched quickly and won't be observable. This leads us to consider endpoints that start close to the horizon (the ``shooting string'' limit), and the strong convergence of \eno{eloss3} for $z_*<z$ leads us to consider the simplest case of keeping only the first, $z_*$-independent term in the expansion. This yields a particularly interesting form of energy loss \cite{ShootingStrings}:
\eqn{eloss4}
{\frac{dE}{dx}=-\frac{\pi}{2}\sqrt{\lambda}T^2\left(\frac{1}{\tilde z_0}+\pi T x\right)^2\,,}
where $\tilde z_0\equiv \pi T z_0 \in [0,1]$. Various limits of this energy loss have interesting physical interpretations: at small $x$, the energy loss looks like a pure $\sim T^2$ energy loss, similar to the pQCD elastic energy loss (with a running coupling); for intermediate $x$, it looks like $\sim x T^3$ with a path dependence similar to the pQCD radiative energy loss; and, finally, for large $x$, it has a novel $\sim x^2 T^4$ behavior. This is an interesting (and a very specific) generalization of the simpler ``abc'' models of energy loss \cite{hg11}, where $dE/dx\propto E^a x^b T^c$. 

A possible way to make our setup more realistic is to add higher derivative $R^2$-corrections to the gravity sector of $AdS_5$, which are the leading $1/N_c$ corrections in the presence of a D7-brane. It has been shown \cite{fng12} that these types of corrections can affect the energy loss significantly and it will be important to explore their effect in the context of finite endpoint momentum strings. We will model the $R^2$-corrections by adding a Gauss-Bonnet term $\delta \mathcal{L}=\lambda_{GB} L^2 /2(R^2-4R_{\mu\nu}^2+R_{\mu\nu\rho\sigma}^2)$ to the five-dimensional action. Here $\lambda_{GB}$ is a dimensionless parameter, constrained by causality \cite{blmsy08} and positive-definiteness of the boundary energy density \cite{hm08} to be $-7/36<\lambda_{GB}\le 9/100$. A black hole solution in this case is known analytically \cite{c02} and hence, using the same procedure as before, we can easily find the energy loss from the finite endpoint momentum in this geometry, solve for the null geodesics and obtain a formula similar to \eno{eloss4} \cite{ShootingStrings}:
\eqn{gb10}
{\frac{dE_{GB}}{dx}=-\sqrt{\lambda}\,T^2 F_n(\lambda_{GB})\left(\frac{G_n(\lambda_{GB})}{\tilde z_0}+\pi T x\right)^2\,.}
Here we have employed a perturbative expansion in $\lambda_{GB}$: functions $F_n$ and $G_n$ are functions of $\lambda_{GB}$ only and do not have a particularly illuminating explicit form, even for small $n$. For $\lambda_{GB}$ as large as $-7/36$, by comparing to the all-order numerical result, it was found that it is enough to go to $n=5$ order in expansion.


\section{Calculation of $R_{AA}$}
\label{confraa}

Before using our proposed energy loss formulas \eno{eloss4} and \eno{gb10} to compute $R_{AA}$ for pions at RHIC and LHC, we will take several steps in order to phenomenologically imitate some of the features of QCD and hence allow for a more realistic computation. First, we will account for roughly three times more degrees of freedom in $\mathcal{N}=4$ SYM than in QCD by relating the temperatures via \cite{g07} $T_{\rm SYM}=3^{-1/4}\,T_{\rm QCD}$ and promote a constant $T_{\rm QCD}$ to a Glauber-like $T_{\rm QCD}(\vec x_\perp,t,\phi)$. To model the transverse expansion of the medium we will use a simple blast wave dilation factor \cite{bg13} $r_{\rm bl}(t)=\sqrt{1+\left(v_T t/R \right)^2}$, where $R$ is the mean nuclear radius and $v_T=0.6$ is the transverse velocity, and replace $\rho_{\rm part}(\vec x_\perp)\to \rho_{\rm part}(\vec x_\perp/r_{\rm bl})/r_{\rm bl}^2$ in the Glauber model. Finally, we use the fragmentation functions \cite{kkp00} to obtain the pionic $R_{AA}$ from the partonic one (neglecting the gluon contribution). 

We use the standard optical Glauber model to compute the participant and binary collisions densities, include the effects of longitudinal expansion and model the spacetime evolution of the temperature. The details of the $R_{AA}$ calculation are given in \cite{ShootingStrings} and consist of finding the initial energy of a jet $p_{T,i}$ using the energy loss formulas \eno{eloss4} and \eno{gb10}, given its final energy $p_{T,f}$, production point $\vec x_\perp$ in the transverse plane and the angle $\phi$ in which it is moving. The nuclear modification factor is then given by a weighted transverse plane average of the ratio of the initial production spectra $d\sigma/dp_T$ (obtained from the LO pQCD CTEQ5 code \cite{WangPrivate}) at final and initial energies.

\begin{figure*}[ht]
\centerline{
\includegraphics[width=0.475\textwidth]{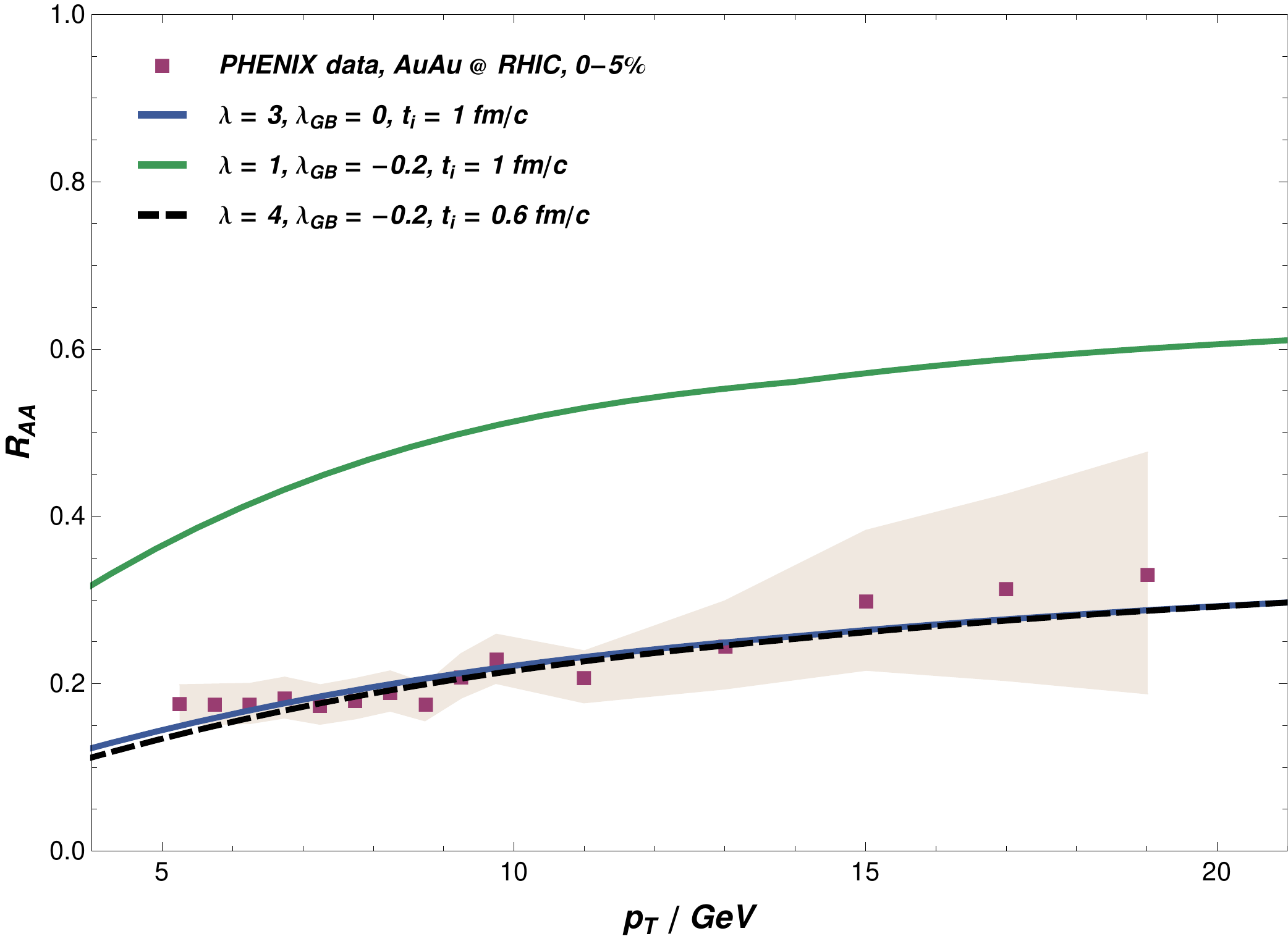}\hskip0.3in
\includegraphics[width=0.475\textwidth]{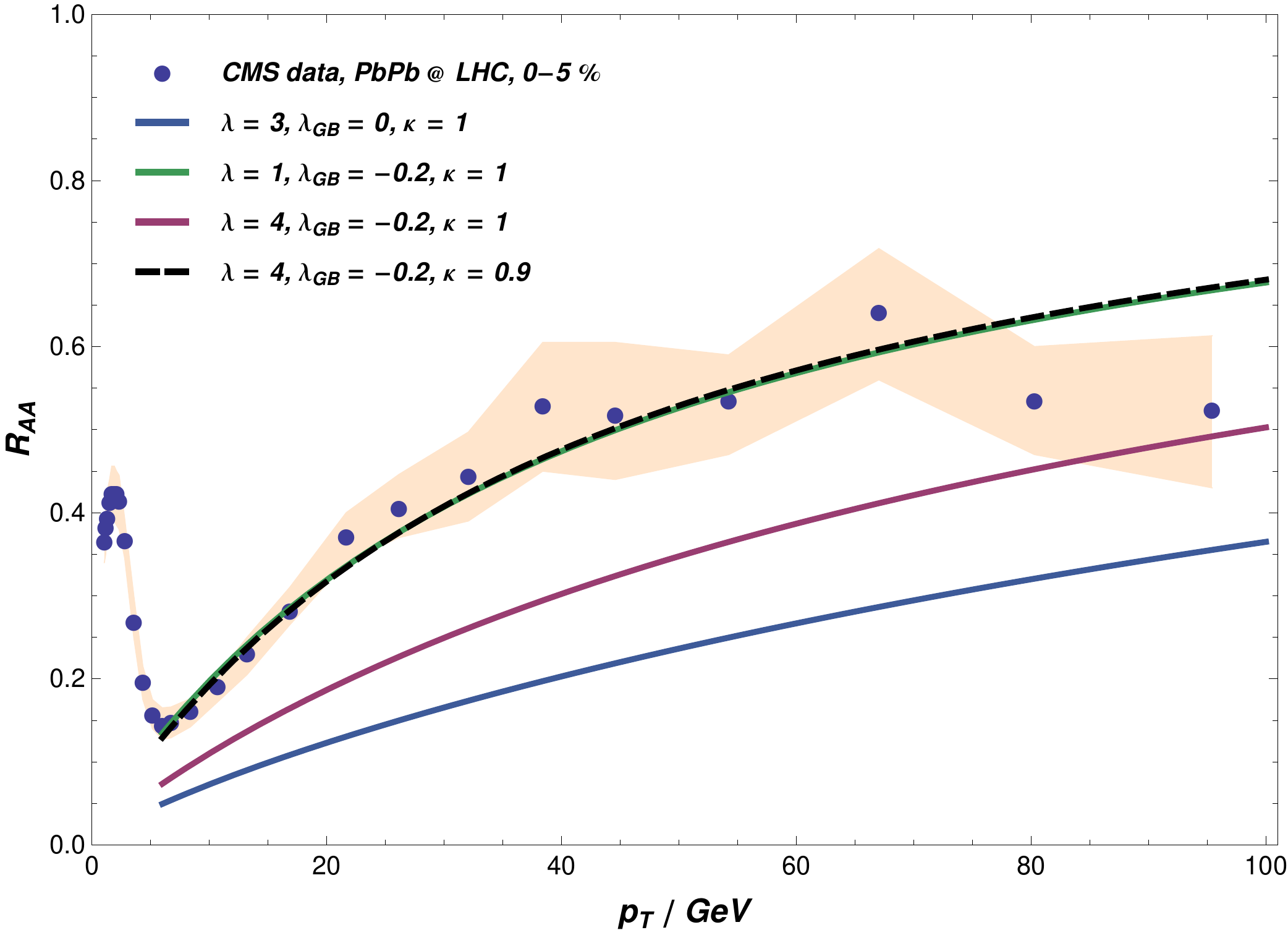}}
\caption{Nuclear modification factor $R_{AA}$ of pions in central collisions at RHIC and LHC. Our calculations are compared to the experimental data from the PHENIX \cite{phenix12} and the CMS \cite{cms12} collaborations for 0-5 \% centrality class. All curves were evaluated with the same impact factor $b=3$ fm, the freezout temperature of $T_{\rm freeze}=170$ MeV, and the initial $\tilde z_0=1$ (from \eno{eloss4}), while the LHC curves also have the same formation time of $t_i=1$ fm/c and an additional temperature adjustment parameter $T\to \kappa T$.}
\label{MainFig}
\end{figure*}

Our main results are shown in Fig. \ref{MainFig}, while the other results can be found in \cite{ShootingStrings}. Qualitatively, our $R_{AA}$ calculations match the data quite well and a good quantitative fit at RHIC, with a reasonable choice of parameters and using \eno{eloss4} (i.e. with no higher derivative corrections), is obtained by choosing $\lambda=3$ (blue curve). In the LHC case, using the same parameters results in a curve (also blue) that is noticeably below the data; this effect is often called the ``surprising transparency'' of the LHC \cite{hg11}, where the effects of temperature increase from RHIC to LHC affect the $R_{AA}$ much more than the competing increase of the production spectra. 

A possible way to obtain a more satisfying fit to LHC is to include the higher derivative corrections via \eno{gb10}. A more phenomenological reason to use these corrections is that they offer a theoretically well defined way to change the shear viscosity, as, in the presence of the Gauss-Bonnet term, $\eta/s=(1-4\lambda_{GB})/(4\pi)$ \cite{blmsy08}. Choosing a maximally negative $\lambda_{GB}=-0.2$ increases the viscosity to $1.8/(4\pi)$, which is, together with our value of the formation time $t_i=1$ fm/c, in the ballpark of the parameters used in some of the most recent hydrodynamic simulations for the LHC \cite{sbh13} necessary to describe the elliptic flow data of light hadrons. This effect puts our curve for $\lambda=1$ on top of the LHC data (green curve), while, as expected, the RHIC data is then over-predicted (also green). 

However, in \cite{sbh13} the initial time $t_i=1$ fm/c used at the LHC was bigger than at RHIC where $t_i=0.6$ fm/c \cite{sbhhs11}, based on the requirements of the hydrodynamic simulations to fit the low $p_T$ elliptic flow data. Using such $t_i$ at RHIC puts us on top of the data for $\lambda=4$ (dashed black curve), while the LHC curve (red) is still somewhat below the data. The reason why lowering the initial time $t_i$ had such a noticeable effect on $R_{AA}$ was because our energy loss formulas \eno{eloss4} and \eno{gb10} have a strong sensitivity to the temperature and in the Glauber model $T\propto t^{-1/3}$. More generally, this strong sensitivity means that a small change in the temperature, $T\to \kappa T$, has the same effect as a large change in the coupling, $\lambda\to\kappa^6\lambda$ or $\kappa^8\lambda$. Hence if there are any phenomenological uncertainties in the effective temperature, such that would allow the LHC temperature to be 10\% lower ($\kappa=0.9$) than given by the simple ratio of the multiplicities, then we can fit the LHC data as well (black dashed curve).


\section{Conclusions}
\label{conclusions}

The framework of finite endpoint momentum strings \cite{fg13} allows for a clear definition of the instantaneous energy loss of light quarks in a strongly coupled SYM plasma, which is identified with the energy flux from the endpoint to the bulk of the string. Using this definition in the case of endpoints that start close to the horizon (``shooting'' strings) leads to a concise and phenomenologically interesting formula \eno{eloss4} \cite{ShootingStrings}. Application of this formula, including the higher derivative $R^2$-corrections (via \eno{gb10}), showed a good independent match with the RHIC and LHC central $R_{AA}$ data for light hadrons. While it is challenging to simultaneously fit both LHC and RHIC data, the choice of $\lambda = 4$ and $\lambda_{GB} = -0.2$ puts our predictions in the ballpark of data {\it provided} we include a $10\%$ reduction of temperature at the LHC relative to straightforward expectations based on multiplicities. Further inclusion of fluctuations and non-conformal effects may provide a simultaneous fit with an even smaller temperature reduction.

\section*{Acknowledgments}
The work of A.F.\ and M.G.\ was supported by U.S. DOE Nuclear Science Grant No. DE-FG02-93ER40764.  The work of S.S.G.\ was supported in part by the Department of Energy under Grant No.~DE-FG02-91ER40671.



\end{document}